\documentclass[sigconf]{acmart}
\usepackage[normalem]{ulem}
\AtBeginDocument{%
  \providecommand\BibTeX{{%
    \normalfont B\kern-0.5em{\scshape i\kern-0.25em b}\kern-0.8em\TeX}}}

\begin{document}

\title{Building Mental Models through Preview of Autopilot Behaviors}

\author{Yuan Shen}
\email{yshen47@illinois.edu}
\affiliation{%
  \institution{University of Illinois at Urbana-Champaign}
  \city{Champaign}
  \state{Illinois}
  \country{USA}
  \postcode{61820}
}

\author{Niviru Wijayaratne}
\email{nnw2@illinois.edu}
\affiliation{%
  \institution{University of Illinois at Urbana-Champaign}
  \city{Champaign}
  \state{Illinois}
  \country{USA}
  \postcode{61820}
}

\author{Katherine Driggs-Campbell}
\email{krdc@illinois.edu}
\affiliation{%
  \institution{University of Illinois at Urbana-Champaign}
  \city{Champaign}
  \state{Illinois}
  \country{USA}
  \postcode{61820}
}



\begin{abstract}
    Effective human-vehicle collaboration requires an appropriate understanding of vehicle behavior for safety and trust. Improving on our prior work by adding a future prediction module, we introduce our framework, called \textit{AutoPreview}, to enable humans to preview autopilot behaviors prior to direct interaction with the vehicle. Previewing autopilot behavior can help to ensure smooth human-vehicle collaboration during the initial exploration stage with the vehicle. To demonstrate its practicality, we conducted a case study on human-vehicle collaboration and built a prototype of our framework with the CARLA simulator. Additionally, we conducted a between-subject control experiment (n=10) to study whether our \textit{AutoPreview} framework can provide a deeper understanding of autopilot behavior compared to direct interaction. Our results suggest that the \textit{AutoPreview} framework does, in fact, help users understand autopilot behavior and develop appropriate mental models.
\end{abstract}




\newcommand{\niv}[1]{\textcolor{red}{#1}}
\maketitle

\section{introduction}
With the recent advance of artificial intelligence, there is increasing use of human-robot collaboration in society, most evidently self-driving vehicles. Autonomous driving systems offer a level of consistency and efficiency that humans cannot match. When coupled with the high cognitive function that humans possess, these systems have tremendous potential. 

To realize the full potential of human-vehicle collaboration, drivers must develop appropriate mental models of their autonomous counterparts. Without a strong understanding of both the system's capabilities and its limitations, drivers can experience unexpected behaviors during operation. Additionally, inappropriate mental models of these systems increase the risk of over-reliance on the vehicle, which gives rise to many dangerous situations. 

We categorized prior research related to human mental model development for robot behavior into two main categories: online interaction~\cite{gao2020joint} or offline presentation of critical states or representative trajectories~\cite{huang2018establishing}. The online approach can be dangerous in the event of unsmooth collaboration, especially when the user is experiencing a new autopilot model (generalizes to both new and current users) for the first time. As for offline interaction, humans are unlikely to spend enough time and effort on critical behavior samples to develop a sufficient mental model, based on the principle of least effort~\cite{zipf2016human}. 
\begin{figure}
\centerline{  \includegraphics[width=0.5\textwidth]{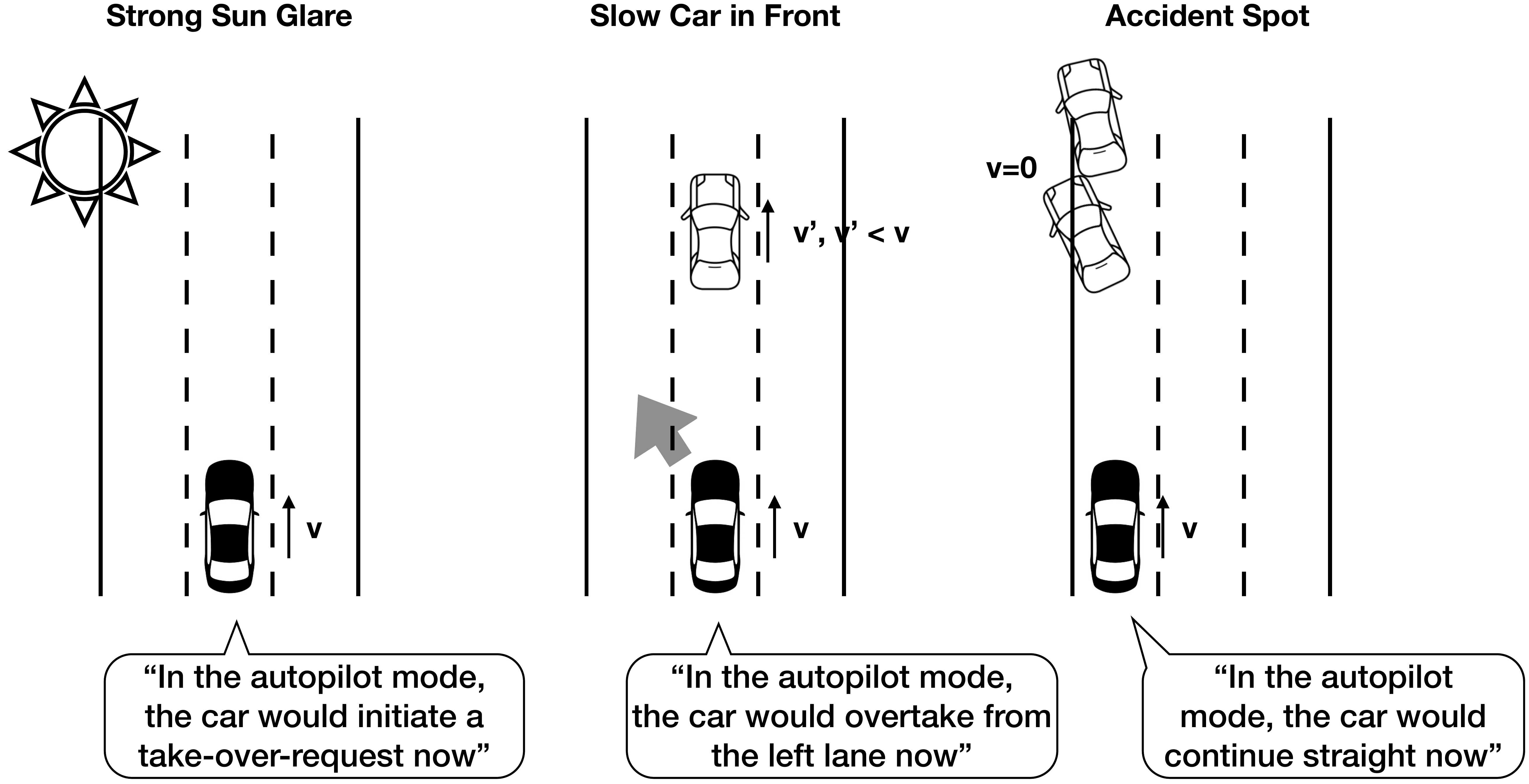}}
\caption{The motivation of our framework. Prior to purchase or deployment of a new autopilot, end-users need to build an accurate mental model of the autopilot's behavior. In particular, drivers should develop an intuition for what scenarios the autopilot will initiate a take-over-request (left figure), when the ego-car would perform a certain action (middle figure), and what scenario the autopilot system might fail in (right figure). Our framework can potentially achieve those purposes through interaction with the driver via a delegate agent which only informs the user about potential action without executing any control on the vehicle.  
\vspace{-0.505cm}}
\label{fig:motivation}
\end{figure}
The purpose of our previously proposed \textit{AutoPreview} framework is to allow humans to preview autopilot behavior interactively. Our objective is to develop tools to help human-collaborators of autonomous vehicles improve their understanding of these systems and establish appropriate levels. Improving on our prior work, we incorporated a future prediction module into the \textit{AutoPreview} framework. We conducted a case study on human-vehicle collaboration to show the practicality of our framework. We plan to generalize this framework for other human-robot collaboration applications in the future.


\section{framework}

\begin{figure}
\centerline{  \includegraphics[width=0.4\textwidth]{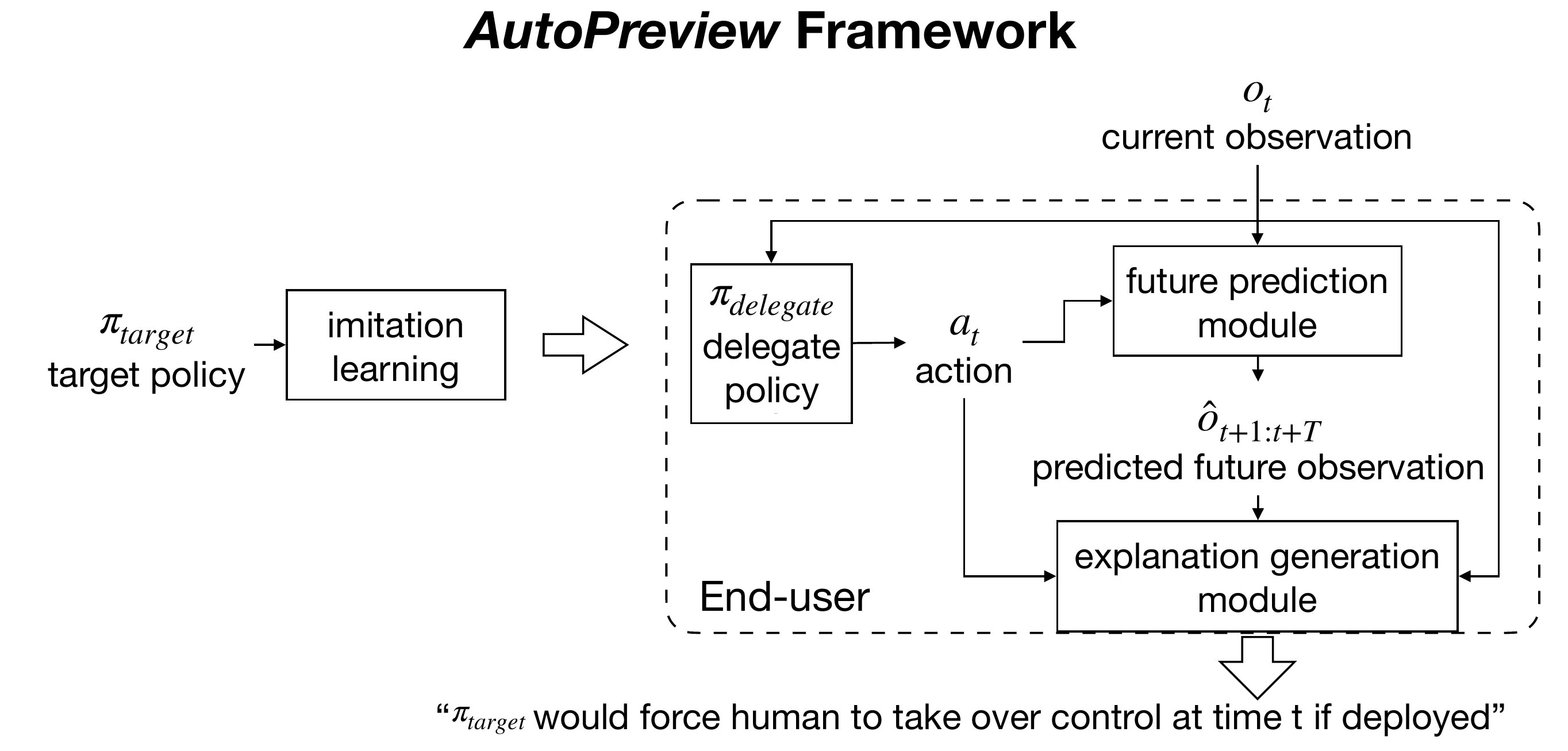}}
\caption{\textit{AutoPreview} Framework. The area enclosed by dotted lines represents the framework logic happening inside the end-user's vehicle. The target autopilot, $\pi_{target}$, refers to the model that people are interested in learning about. The delegate autopilot, $\pi_{delegate}$, is a model whose behavior matches that of $\pi_{target}$ and informs human drivers about the potential actions of $\pi_{target}$ based on the current driving state. The future prediction module generates the potential future implications of taking a specific action. \vspace{-0.505cm}}
\label{fig:framework}
\end{figure}
As illustrated in Figure \ref{fig:motivation}, the motivation of \textit{AutoPreview} is to provide an easy-to-use and safe tool for potential users to understand, evaluate, and develop appropriate mental models of autopilot behavior. We aim to enable end-users to preview the behaviors of the target autopilot, $\pi_{target}$, indirectly through a delegate autopilot, $\pi_{delegate}$ (Figure \ref{fig:framework}). In our prototype, we directly use the target autopilot as the delegate autopilot to test the effectiveness of this framework in the ideal context ie. a delegate autopilot that perfectly matches the behavior of the target autopilot. Building on our prior work, ~\cite{10.1145/3411763.3451591}, we employ a prediction module which will predict potential future failures of the vehicle, $o_{t+1:t + T}$ as a result of action, $a_t$ (Figure \ref{fig:framework}). In our prototype, we extracted the future effect by estimating collisions based on lateral and longitudinal acceleration and distance. Note that $\pi_{delegate}$ will not execute any control actions over the vehicle, rather it will feed in potential actions and future observations to the explanation generation module, which will then prepare a visual or verbal explanation for the user. Under the \textit{AutoPreview} framework, drivers can manually control their vehicle to actively learn from interesting scenarios and can evaluate the newly released autopilot under those conditions~\cite{felder2009active}. 
\section{experiment}
To show the practicality of the \textit{AutoPreview} framework, we did a case study on the human-vehicle collaboration task in the CARLA simulator~\cite{dosovitskiy2017carla}. In particular, we conducted a between-subject online control experiment with two conditions (n=5): directly observing target policy and observing through \textit{AutoPreview}. Ten participants between the ages of 18 and 30 participated in our study voluntarily. Our experiment hypothesis is that \textit{AutoPreview} can establish at least the same level of understanding of the target autopilot's lane-switching behavior than that formed through direct observation of the target policy.

\begin{figure}
\centerline{  \includegraphics[width=0.5\textwidth]{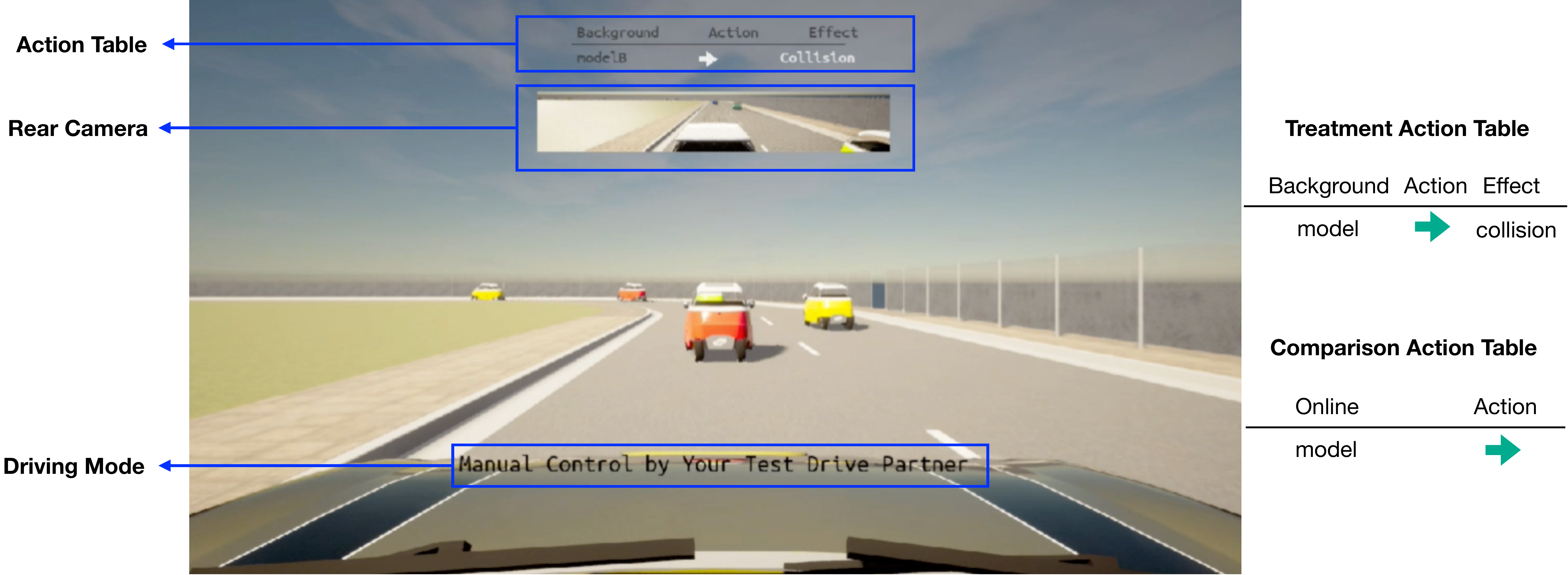}}
\caption{Our framework prototype. The left screenshot shows the driver-perspective video interface that is used in our control experiment. For the treatment group (upper-right), the action column describes the potential actions based on the delegate autopilot's output and the potential future effect of this action. For the comparison group (lower-right), the arrow icons in the action column reflect the vehicle's lane-changing operation, which is directly controlled by the target autopilot. \vspace{-0.54cm}}
\label{fig:interface}
\end{figure}

We used a Model-Predictive-Control controller as the target autopilot policy and customized our CARLA map such that the only interesting target policy behavior was the timing of the car's lane-switching actions. To simulate the driving experience, we prepared 3-minute videos for different condition groups and asked our participants to learn about the target policy behavior through their respective videos (Figure \ref{fig:interface}). In the post-experiment stage, participants were told to specify the timestamp that the target policy would most likely perform a lane-switching operation, in eight different 5-second test scenarios. We measured the users' level of understanding about target policy behavior in terms of the average L1 error between the user predicted and ground-truth lane-switching timestamp across test scenarios, weighted by their reported prediction confidence.

The five participants who used \textit{AutoPreview} (M = 0.67, Min=0.29, Max=0.96, SD = 0.27) compared to the five participants in the control group (M = 1.09, Min=0.74, Max=1.55, SD = 0.35) demonstrated significantly less weighted timing error (t(8) = 2.69, p < 0.05, Hedges' g = 1.34). This suggests that the \textit{AutoPreview} method can enable more accurate action timing prediction for target autopilot behavior. However, the reported participant confidence in the treatment group (M=0.55, SD=0.10) was less than that in the comparison group (M=0.78, SD=0.17)) (one-tailed Mann-Whitney U, U(5,5)=23.5, p < 0.05) indicating that the notification system for this framework requires further research.

\section{conclusion}
We propose the \textit{AutoPreview} framework, which abstracts autopilot policies into explainable policies for viewing and exploring online. Our preliminary findings suggest that the \textit{AutoPreview} method is intuitive and can help users understand autopilot behavior in terms of exact action timing prediction. 

There are several limitations with this framework. First, the delegate autopilot can potentially report actions in states that the target autopilot is unlikely to visit, since it presents actions based on current observations without considering the state visitation frequency. Moreover, our prototype can only report the action triggering moment related information to drivers. Subtle behavior, e.g., how soft the brake would be, requires future research to explore. 

Despite its limitations, the \textit{AutoPreview} framework shows promise for mental model development in the autonomous driving domain.

\bibliographystyle{ACM-Reference-Format}
\bibliography{sample-base}
\end{document}